%%%%%%%%%%%%%%%%%%%%%%%%%%%%%%%%%%%%%%%%%%%%%%%%%%%%%%%%%%%%%%%%%%%%%%%%%%
%                                                                        %
%                                                                        % 
%        by J. Bouttier, P. Di Francesco and  E. Guitter                 %
%                TEX file, using harvmac.tex macros                      %
%                                                                        %
%                                                                        %
%%%%%%%%%%%%%%%%%%%%%%%%%%%%%%%%%%%%%%%%%%%%%%%%%%%%%%%%%%%%%%%%%%%%%%%%%%
\input harvmac 
\input epsf.tex

\overfullrule=0mm

\newcount\figno
\figno=0
\def\fig#1#2#3{
\par\begingroup\parindent=0pt\leftskip=1cm\rightskip=1cm\parindent=0pt
\baselineskip=11pt
\global\advance\figno by 1
\midinsert
\epsfxsize=#3
\centerline{\epsfbox{#2}}
\vskip 12pt
{\bf Fig. \the\figno:} #1\par
\endinsert\endgroup\par
}
\def\figlabel#1{\xdef#1{\the\figno}}
\def\encadremath#1{\vbox{\hrule\hbox{\vrule\kern8pt\vbox{\kern8pt
\hbox{$\displaystyle #1$}\kern8pt}
\kern8pt\vrule}\hrule}}

%Macros 
%%%%%%%%%%%%%%%%%%%%%%%%%%%%%%%%%%%%%%%%%%%%%%%%%%%%%%%%%%%%%%%%%

\def\IR{\relax{\rm I\kern-.18em R}}
\font\cmss=cmss10 \font\cmsss=cmss10 at 7pt

\font\cmss=cmss10 \font\cmsss=cmss10 at 7pt
\def\IZ{\relax\ifmmode\mathchoice
{\hbox{\cmss Z\kern-.4em Z}}{\hbox{\cmss Z\kern-.4em Z}}
{\lower.9pt\hbox{\cmsss Z\kern-.4em Z}}
{\lower1.2pt\hbox{\cmsss Z\kern-.4em Z}}\else{\cmss Z\kern-.4em Z}\fi}
\def\IN{\relax{\rm I\kern-.18em N}}

%%%%%%%%%%%%%%%%%%%%%%%%%%%%%%%%%%%%%%%%%%%%%%%%%%%%%%%%%%%%%%%%%

\Title{\vbox{\hsize=3.truecm \hbox{SPhT/02-075}}}
{{\vbox {
%\centerline{}
\bigskip
\centerline{Counting Colored Random Triangulations}
}}}
\bigskip
\centerline{J. Bouttier\foot{bouttier@spht.saclay.cea.fr}, 
P. Di Francesco\foot{philippe@spht.saclay.cea.fr} and
E. Guitter\foot{guitter@spht.saclay.cea.fr}}
\medskip
\centerline{ \it Service de Physique Th\'eorique, CEA/DSM/SPhT}
\centerline{ \it Unit\'e de recherche associ\'ee au CNRS}
\centerline{ \it CEA/Saclay}
\centerline{ \it 91191 Gif sur Yvette Cedex, France}
\bigskip
%\vskip .5in
%abstract
\noindent 
We revisit the problem of enumeration of vertex-tricolored 
planar random triangulations solved in [Nucl. Phys. B 516 [FS]
(1998) 543-587] in the light of recent combinatorial
developments relating classical planar graph counting problems
to the enumeration of decorated trees. We give a direct
combinatorial derivation of the associated counting function,
involving tricolored trees. This is generalized to arbitrary 
$k$-gonal tessellations with cyclic colorings and checked by
use of matrix models.

%\draft
\Date{06/02}

%references
\nref\DEG{P. Di Francesco, B. Eynard and E. Guitter,
{\it Coloring Random Triangulations},
Nucl. Phys. {\bf B516 [FS]} (1998) 543-587.}
\nref\TUT{W. Tutte, {\it A Census of Planar Maps}, Canad. Jour. of Math. 
{\bf 15} (1963) 249-271.}
\nref\SCH{G. Schaeffer,  {\it Bijective census and random 
generation of Eulerian planar maps}, Electronic
Journal of Combinatorics, vol. {\bf 4} (1997) R20}
\nref\BMS{M. Bousquet-M\'elou and G. Schaeffer,
{\it Enumeration of planar constellations}, Adv. in Applied Math.,
{\bf 24} (2000) 337-368.}
\nref\KONTS{After M. Kontsevitch, Mathematical Intelligencer,
Vol. {\bf 19}, No. 4 (1997) 48.}
\nref\PS{D. Poulhalon and G. Schaeffer, 
{\it A note on bipartite Eulerian planar maps}, preprint (2002),
available at {\sl http://www.loria.fr/$\sim$schaeffe/}}
\nref\BIPZ{E. Br\'ezin, C. Itzykson, G. Parisi and J.-B. Zuber, {\it Planar
Diagrams}, Comm. Math. Phys. {\bf 59} (1978) 35-51.}
\nref\DGZ{P. Di Francesco, P. Ginsparg
and J. Zinn--Justin, {\it 2D Gravity and Random Matrices},
Physics Reports {\bf 254} (1995) 1-131.}
\nref\EY{B. Eynard, {\it Random Matrices}, Saclay Lecture Notes (2000),
%\hfill\break% 
available at {\sl http://www-spht.cea.fr/lectures\_notes.shtml} }

%text
\newsec{Introduction}

Graph-coloring problems are a classical subject of combinatorics, where 
the celebrated 4-color theorem is probably the most famous result.
Parallely the statistics of randomly generated colored graphs 
play an important role in physics
in connection to various statistical models.

Among the coloring problems, a fundamental one is that
of generating and counting {\it vertex-tricolored random triangulations},
a task completed recently in Ref. \DEG, where a generating
function for tricolored triangulations was obtained using matrix model
techniques. This generalized an earlier standard result by W. Tutte
where these triangulations were counted without colors \TUT. 
 
The scope of the present paper is to revisit this problem 
of counting vertex-tricolored random triangulations in the light 
of new combinatorial developments
relating various problems of graph enumeration to that, much simpler,
of counting decorated trees [\xref\SCH,\xref\BMS].

This note is organized as follows. We first recall in Section 2 
the counting problem and Tutte's result which we re-derive in
a more general form in Appendix A by use of a matrix integral.
We also recall the formula for the counting function with colors
first derived in Ref. \DEG. In Section 3, 
we present a purely combinatorial proof of this formula  
using techniques borrowed from Refs. [\xref\SCH,\xref\BMS]. 
Section 4 gathers a few concluding remarks
as well as some natural generalizations to colored $k$-gonal
tessellations, corroborated by the matrix integral results
of Appendix A.

\newsec{Generalities}

\subsec{The Problem}

\fig{An example of rooted vertex-tricolored triangulation. 
We indicate the colors $1,2,3$ and the natural orientation of edges 
$1\to 2\to 3\to 1$. The root is the (12) edge marked with a white 
arrow. The graph is clearly Eulerian as orientations of edges 
alternate around each vertex.}{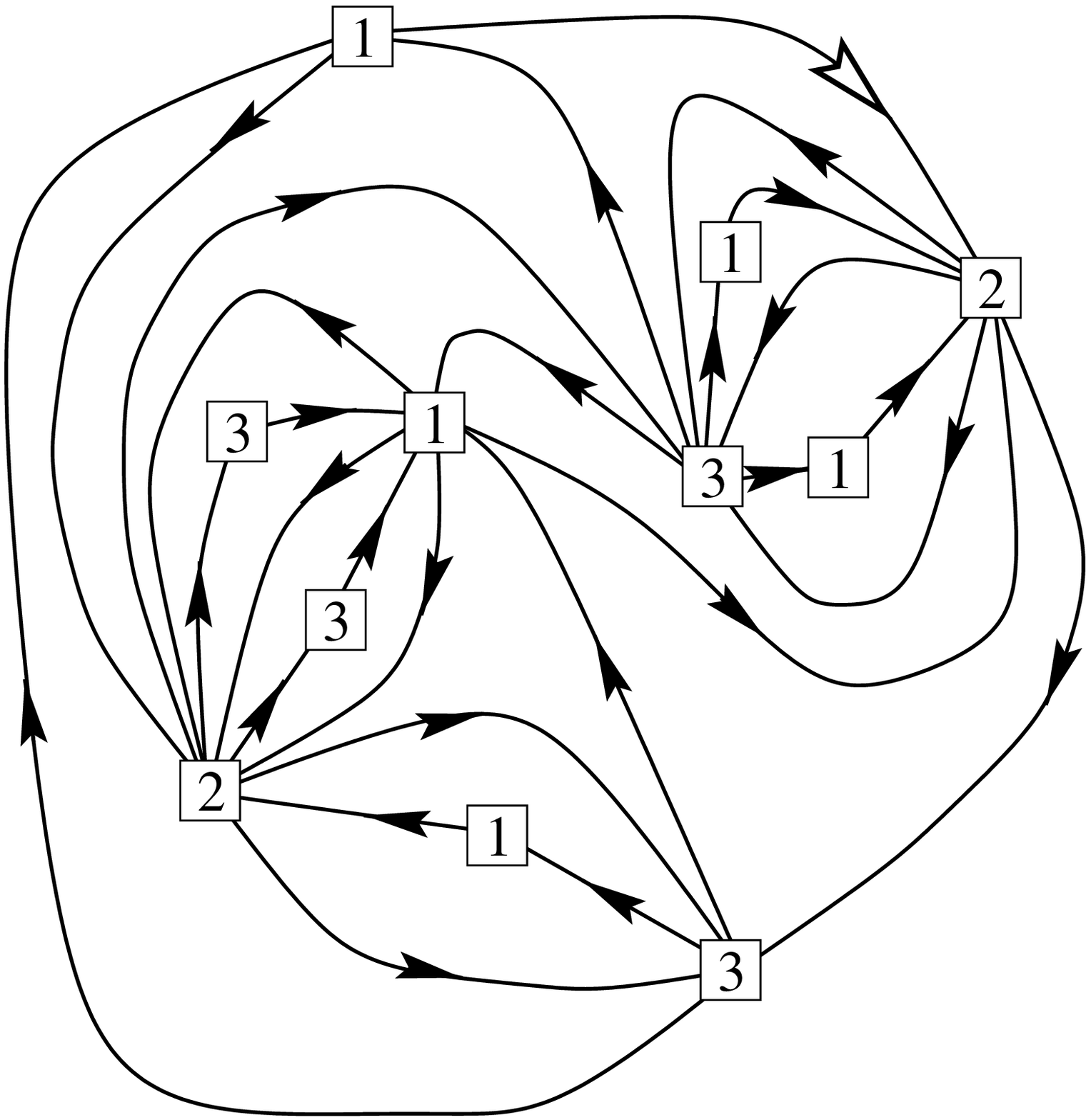}{8.truecm}
\figlabel\trianBW

We wish to generate so-called ``planar random triangulations" namely arbitrary
triangulations of the sphere, with vertices colored using three distinct
colors say 1,2,3 in such a way that no two adjacent vertices be of the same 
color (see Fig. \trianBW). 
To avoid spurious symmetry factors, we furthermore decide to mark and orient
one edge of the triangulation and demand that it points from a vertex of color
1 to one of color 2. Finally, we attach a weight $t$ per edge\foot{
Note that the number of edges
colored (12) is the same as that colored (23) and that colored (13), and is 
also half the number of triangles in the triangulation.} colored (12)
and a weight $x_i$ per vertex of color $i\in \{1,2,3\}$. 

It is easy to see that for a given triangulation, if such a tricoloring exists,
then it is unique. This is to be contrasted
to the problem of edge-tricoloring of random planar triangulations where 
a given edge-tricolorable triangulation may be colored in many ways. 
In Ref. \DEG\ it was shown that the vertex-tricolorability of a triangulation 
is equivalent to its two-dimensional foldability, i.e.
to the possibility of making all its triangles equilateral by
folding it onto the plane. This property
holds for arbitrary genus of the triangulated surface. 

On the other hand, for {\it planar} (genus zero) graphs, a necessary and sufficient 
condition for a triangulation to be vertex-tricolorable is that it be
Eulerian, namely that each vertex have an even valency. This condition is also
equivalent to the {\it face-bicolorability} of the triangulation, where
the two colors of the faces correspond respectively to a clockwise or 
counterclockwise cyclic order of the vertex colors around them.
That the condition of face-bicolorability be sufficient
to ensure vertex-tricolorability is non-trivial (see e.g. Ref. \KONTS), and holds only in 
genus zero. 

\subsec{Vertex-tricolorable triangulations}

For $x_1=x_2=x_3=1$, the counting problem reduces to finding the number of 
vertex-tricolorable planar triangulations or
equivalently of planar Eulerian triangulations with $2n$ faces and a marked
oriented edge. This was done by W. Tutte in the dual language 
under the name of ``rooted bicubic maps", i.e. graphs which are planar, rooted,
trivalent, and bipartite, with the result \TUT:
\eqn\tutt{ e_n = {3\ .\  2^{n-1} \over (n+1)(n+2)}\  {2n \choose n}  }
The corresponding generating function $E(t)$ reads
\eqn\gentut{ E(t)={1\over 32 t^2} \big((1-8t)^{3\over 2} -1+ 12 t - 24 t^2\big) }  
Beyond the original combinatorial proof of Ref. \TUT, this result may be easily recovered
and extended to bipartite $k$-valent planar graphs  
using a two-matrix integral (see Appendix A for details). 

\subsec{Vertex-tricolored triangulations}

Keeping track of the numbers of vertices of each color, i.e. having $x_1,x_2,x_3$
arbitrary is another story.
In terms of matrix integrals, it leads to a much more involved problem, which was solved
in Ref. \DEG. In this formulation, the triangulations themselves (not their duals)
are generated in two steps: one first generates arbitrary vertex-bicolored graphs
with weights $x_1,x_2$ for the first two colors, by use of a two-matrix model with logarithmic 
potential;  
the vertices of the third color $3$ are added in the center of each face of the above graphs,
and the weight $x_3$ simply arises from the size $n= N x_3$ of the matrices, where $N$
is a coefficient in the potential.
 
Remarkably enough, the result of Ref. \DEG\ for the generating function $E(t;x_1,x_2,x_3)$
takes the simple form:
\eqn\lutinit{ \eqalign{ E(t;x_1,x_2,x_3) &=
{U_1U_2U_3\over t^2} (1-U_1-U_2-U_3) \cr
U_1 (1-U_2-U_3) &= t x_1\cr
U_2 (1-U_3-U_1) &= t x_2\cr
U_3 (1-U_1-U_2) &= t x_3\cr}}
where $U_i\equiv U_i(t;x_1,x_2,x_3)$ satisfy $U_i(0;x_1,x_2,x_3)=0$.
It was also found in Ref. \DEG\ that this can be differentiated into 
$dE/dt=U_1U_2U_3/t^3$. Note finally that
when $x_1=x_2=x_3=1$, we have $U_1=U_2=U_3=U(t)=(1-\sqrt{1-8t})/4$ 
and \lutinit\ reduces to \gentut.

\newsec{Combinatorial Solution}

The simplicity of Tutte's result for $e_n$ \tutt\ clearly related to Catalan numbers
suggests a connection to the counting of (possibly decorated) trees.
It was shown recently \PS\ that a one-to-one correspondence indeed exists
between Eulerian triangulations and so called ``well-balanced trees" which allows
for an alternative proof of \gentut. This is a particular example of a more
general correspondence between so-called ``constellations" and 
suitable trees \BMS.  

The generalization to the colored case is also very suggestive, as the
$U_i$'s are reminiscent of generating functions of decorated rooted trees.  
As we shall see now, the construction of Ref. \BMS,
when dressed with colors, leads to a combinatorial proof of \lutinit.

\subsec{Conjugated trees}

\fig{A rooted blossom binary tree. It is generically obtained by first
drawing a rooted binary tree (with white inner vertices and white leaves,
including the root on top)
and adding in the middle of each inner edge a black vertex with a 
bud pointing to either side of the edge.}{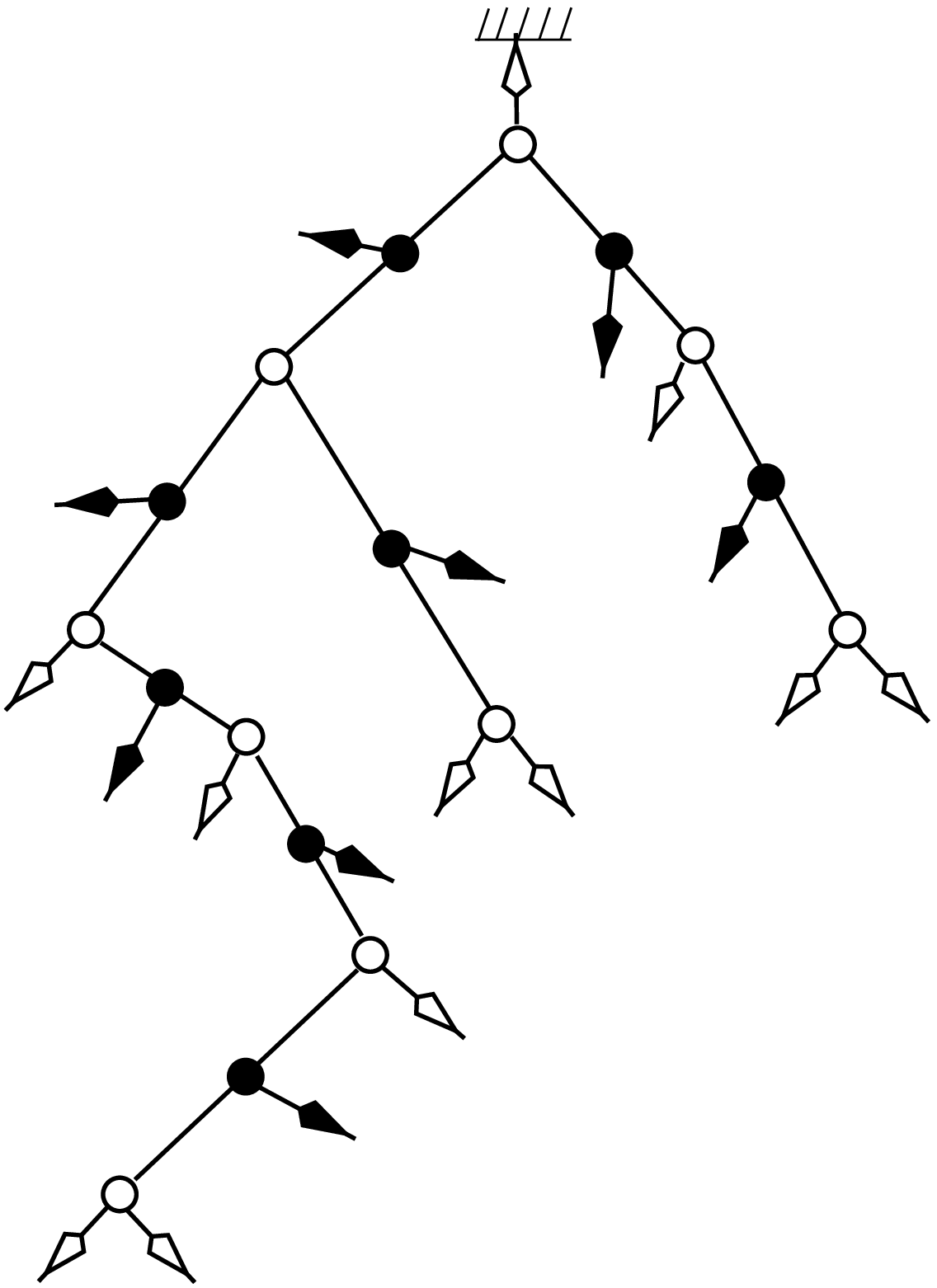}{5.truecm}
\figlabel\blossomBW

Let us first recall the construction of Ref. \BMS\ establishing a
one-to-one correspondence between on one side
Eulerian triangulations with a marked oriented edge, and on the
other side well-balanced blossom binary trees defined as follows.
One starts with a rooted binary tree with $n$ inner (white) vertices, $n-1$ inner edges,
$n+2$ (white) leaves (including the root). There are
\eqn\bintree{ c_n= {1\over n+1} {2n \choose n}}
such trees, where $c_n$ denotes the $n$th Catalan number.
These trees are decorated by adding buds to each inner edge in one of the two
possible ways depicted in Fig. \blossomBW. 
This adds $n-1$ (black) vertices and $n-1$
(black) buds to the trees, now called blossom trees. 
There are $2^{n-1} c_n$ such blossom trees.

\fig{The matching of buds and leaves of a rooted blossom binary tree. We are 
left with three unmatched leaves. The tree is well-balanced iff the root
is one of them, as is the case here.}{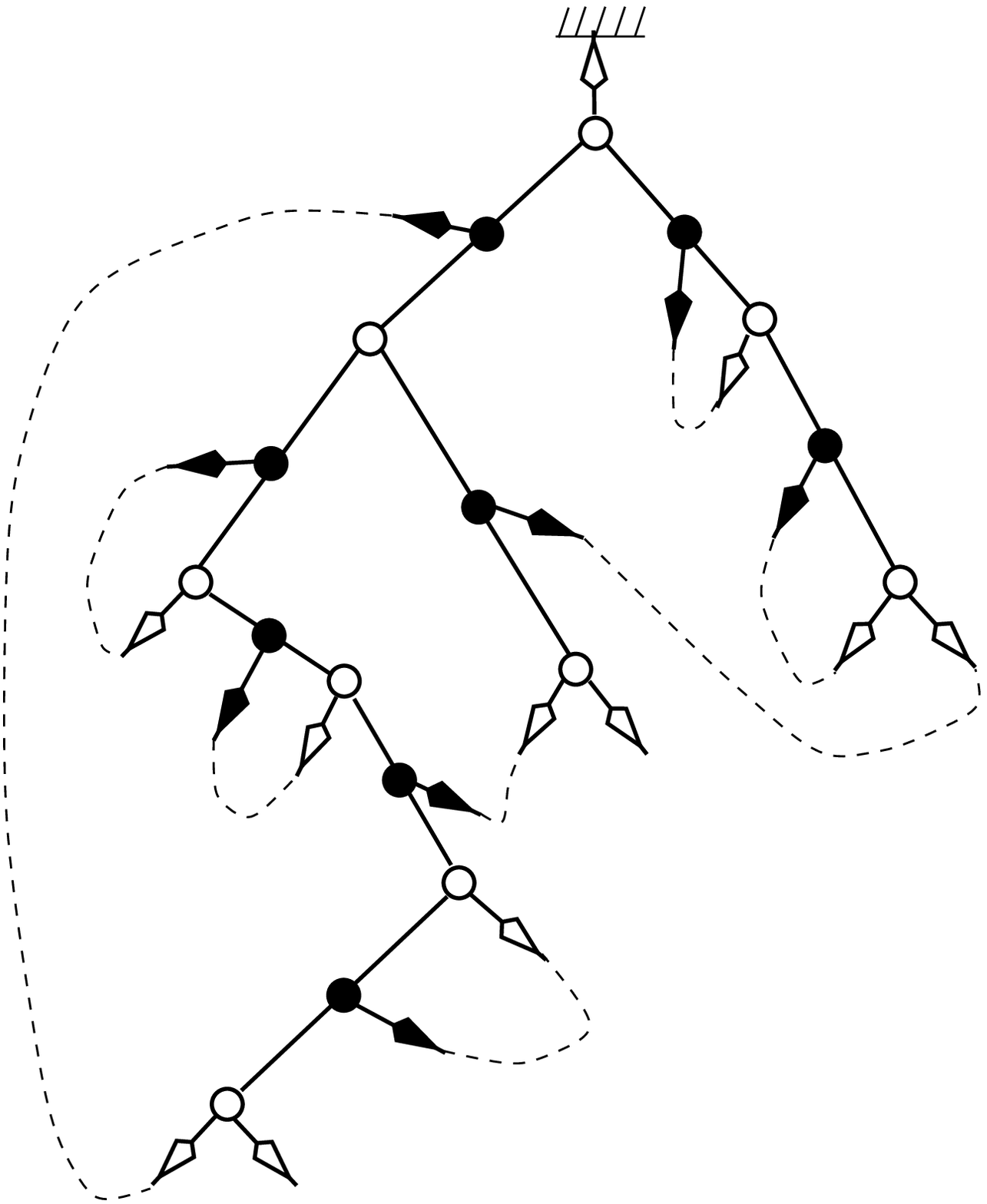}{5.5truecm}
\figlabel\wellbaBW
\fig{Starting from the well-balanced blossom tree of Fig. \wellbaBW,
the three unmatched leaves are connected into an additional black vertex (a)
and the fusing of all bud-leaf pairs into edges leads to a bipartite 
trivalent graph (b). The latter is rooted at the edge pointing from the 
added vertex to the root of the tree. Coloring with color 1 (resp. 2) 
the face on the right (resp. left) of this edge induces a unique tricoloring
of all faces, as indicated. 
This graph is dual to the vertex-tricolored triangulation of Fig.
\trianBW.}{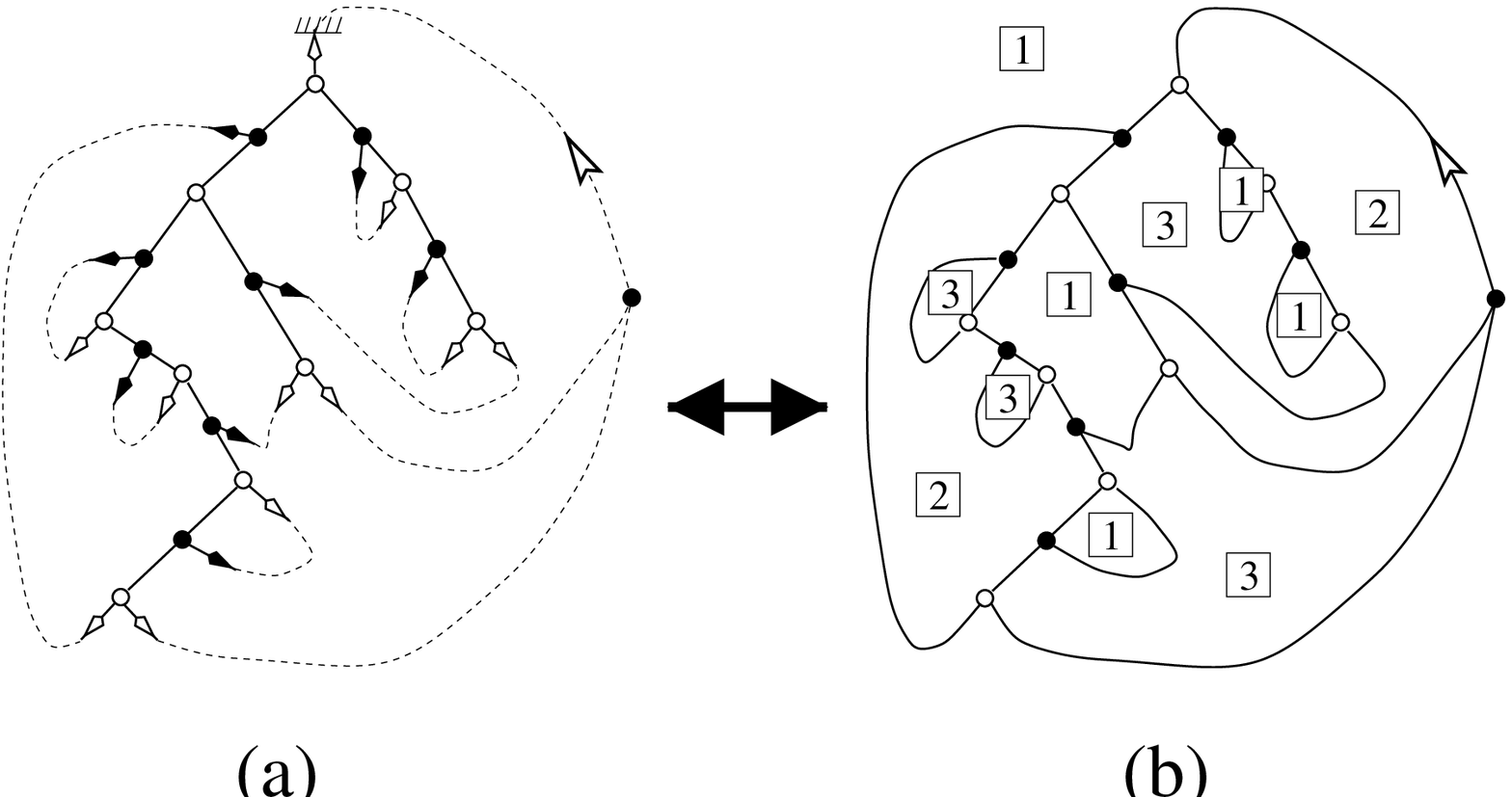}{14.truecm}
\figlabel\evolBW

Black buds are then linked to white leaves as shown in Fig. \wellbaBW\ 
by matching them iteratively to the closest available leaf
in counterclockwise direction in such a way that the resulting graph
is planar (no intersection of edges). This leaves exactly three 
white leaves unmatched among the $n+2$ original ones, separating three arch systems
of paired bud-leaves. 
The blossom tree is said to be well-balanced if the root is one of
these three unmatched white leaves. The number of such objects
reads \BMS\
\eqn\welba{ {3 \over n+2}\  \times \ 2^{n-1} c_n = \ e_n }   
To any well-balanced blossom binary tree 
one associates a rooted bipartite trivalent planar graph (rooted bicubic map)
by fusing each bud-leaf pair into an edge and  connecting the three
unmatched leaves to an additional (black) vertex (see Fig. \evolBW). 
The marked edge is that pointing
from this vertex to the root of the tree. The desired Eulerian triangulation is 
the dual of this rooted graph, with a marked edge obtained by
the Amp\`ere rule. This triangulation has $2n$ faces ($n$ white and $n$ black), 
$3n$ edges and $n+2$ vertices (by Euler's relation).  

\fig{The covering forest of the triangulation of Fig. \trianBW, made of
the three trees obtained as follows: the three edge of
the face to the left of the rooted edge serve as roots for the trees
and point to three origins labeled by $0$. Each vertex is labeled by
its minimal geodesic ``distance" to the nearest origin. 
The {\it leftmost} minimal paths linking origins to vertices have been 
thickened and form the three trees.}{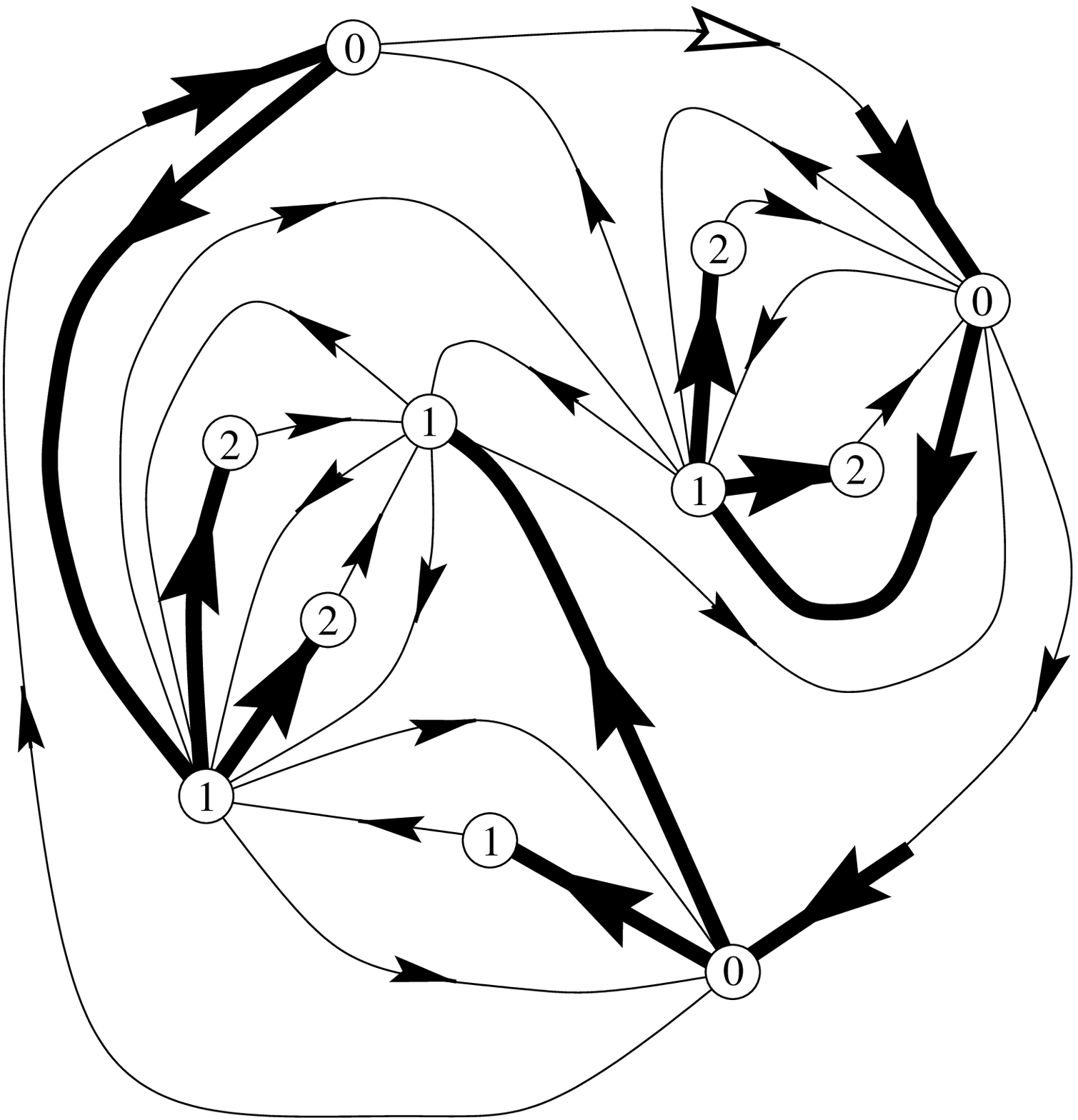}{8.truecm}
\figlabel\treesBW

Conversely, starting from a rooted Eulerian triangulation,
a well-balanced blossom binary tree is constructed from the
trivalent bipartite graph dual to the triangulation by removing one vertex
and cutting $n+2$ edges as follows. 
The orientation of the marked edge in the triangulation induces an orientation
of all the edges by imposing that the orientations around a triangle
be either clockwise or counterclockwise. 
This orientation is equivalent to a bicoloration of the triangles (white for clockwise,
black for counterclockwise) and, upon adding colors, 
also corresponds to the cyclic order 
of the three vertex colors around the triangles (namely $1,2,3$ or $1,3,2$).
One first opens the triangle  
adjacent to the marked oriented edge on its left, so as to obtain three root edges pointing 
toward three vertices with distinct colors which we will call ``origins".
To each vertex of the triangulation one associates its geodesic ``distance" from these origins 
defined as the minimal length of all orientation-respecting paths from any of the origins 
to this vertex. As a consequence of the vertex-tricolorability, all minimal paths 
to a given vertex originate from the same origin. Among these paths, one can 
moreover select the {\it leftmost} one \BMS. 
The edges of all the leftmost minimal paths form a ``covering forest"
made of three disjoint trees 
rooted at the origins and covering all vertices of the graph,
as shown in Fig. \treesBW. The
duals of these edges are those to be cut on the dual graph 
in order to get the desired well-balanced tree,  
obtained by moreover erasing the vertex dual to the opened face. 
It is easily seen that this indeed produces the inverse of the former construction 
as the above three trees are dual to the three arch systems formed by the
bud-leaf pairings and delimited by the three unmatched leaves 
(see Ref. \BMS\ for a detailed proof). 

This mapping allows to identify $E(t)$ \gentut\ as the generating function
of the numbers of well-balanced rooted binary blossom trees with a weight $t$
per inner white vertex. 

\subsec{Colors}

\fig{Color assignment for leaves and half-edges of the blossom tree
of Fig. \blossomBW. The leaves are in one-to-one correspondence with
the faces of the trivalent bipartite graph, dual to the vertices of the 
original triangulation, and receive a weight $x_i$
according to their color, here indicated within thick squares.
Half-edges (i.e. links from black to white vertices) are colored
according to the face on their left, as indicated by flags, but
receive no additional weight. Buds also receive a color 
(not indicated here) identical to that of the connected leaf. Colors
appear necessarily in clockwise (resp. counterclockwise) cyclic order around
white (resp. black) vertices.}{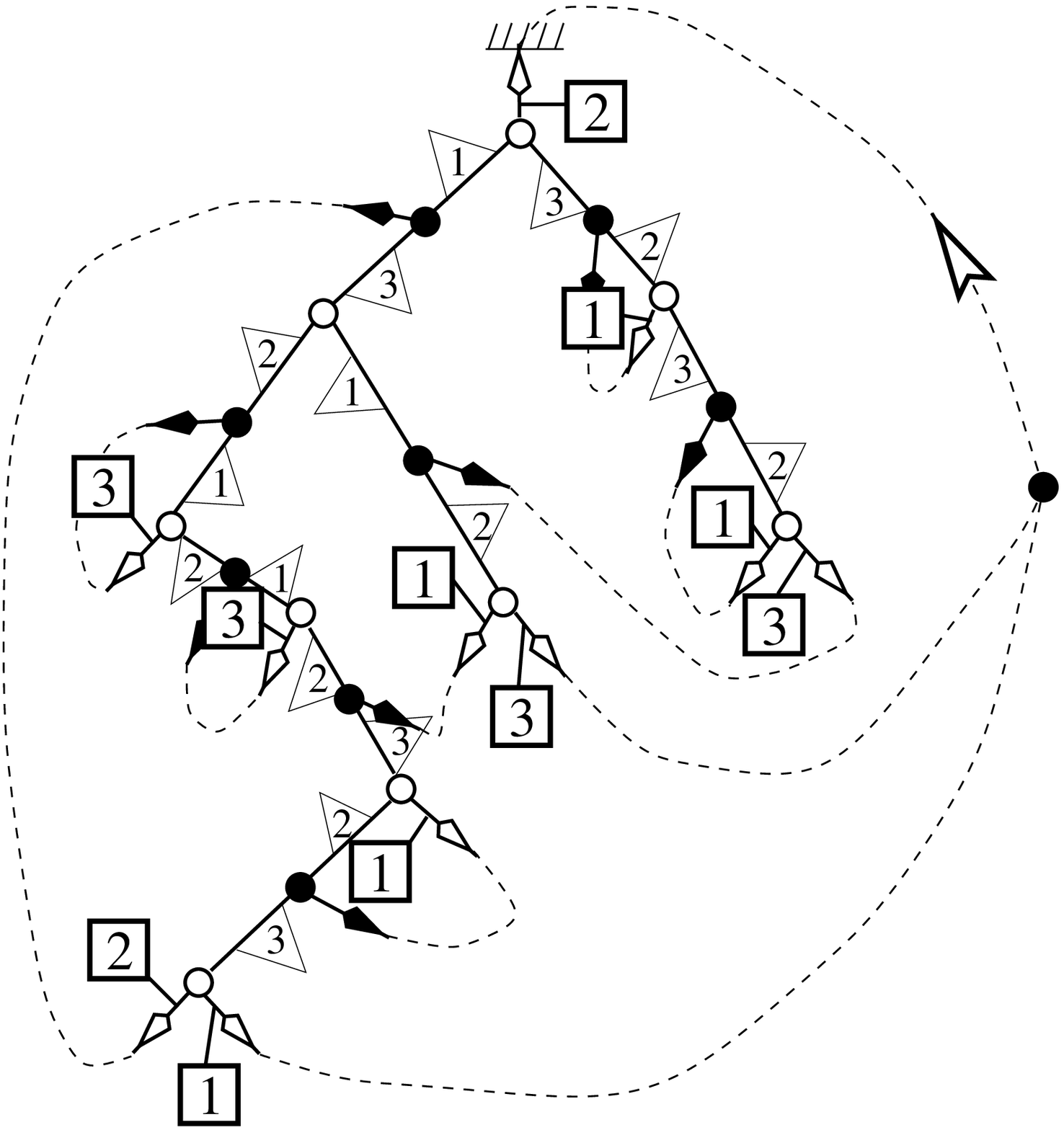}{7.5truecm}
\figlabel\colors

It is now quite simple to reinstate the color weights as the above construction
allows to keep track of the tricoloration. Indeed, the $n+2$ vertices of the triangulation
are in one-to-one correspondence with the $n+2$ white leaves of the well-balanced
tree. To any vertex, we first associate the incident edge of the above-constructed
covering forest pointing to it
(to a root vertex, we associate the corresponding root edge). 
The dual of this edge is cut in the above procedure 
hence associated with a unique (white) leaf. We decide to transfer the color weight 
$x_i$ from the original vertex of the triangulation to this leaf of the associated
well-balanced tree (see Fig. \colors).

Beside this coloring of the leaves, the original coloring of the triangulation
induces a coloring of the whole well-balanced tree as follows.
We first attach a color to each {\it edge} of the trivalent bipartite graph
dual to the triangulation. The edges of this graph are naturally oriented
from black to white vertices and its faces inherit the colors $1,2$ or $3$
of the dual vertices. We decide to attach to each oriented edge the color of the
face immediately to its left. Note that edge colors around a white (resp. black)
vertex necessarily appear in clockwise (resp. counterclockwise) cyclic order.
Keeping this edge-coloring assignment, we may repeat the above construction of
a well-balanced blossom binary tree, whose half-edges, buds and leaves are now colored 
accordingly (see Fig. \colors). In particular, for leaves, 
the color inherited from the associated cut edge coincides with that 
assigned above when transferring colors from vertices to leaves. 

Note that, on the well-balanced tree, the coloring is uniquely fixed by demanding that 
the three colors adjacent to white (resp. black vertices) be in clockwise 
(resp. counterclockwise) cyclic order, and that the root half-edge be of color $2$
(with our conventions).  Such a tree will be called well-colored. 
Note also that only leaves receive a weight $x_i$ according to their color. 
This in turn allows to identify $E(t;x_1,x_2,x_3)$ of \lutinit\ as the
generating function of the well-colored well-balanced rooted blossom binary trees
(with a root edge of color $2$) with a weight
$x_i$ per leaf of color $i$ (including the weight $x_2$ for the root).

\subsec{Generating functions}

Using the above equivalence, the formula \lutinit\ for $E(t;x_1,x_2,x_3)$
can now be obtained thanks to the following trick.
We first compute the generating function $dE/dt$ incorporating
an extra marking of a white {\it vertex} (the weight $t$ per edge (12) 
is equivalently a weight per white face of the triangulation,
hence per white vertex of the well-balanced tree). Erasing
this vertex cuts out the colored rooted well-balanced binary blossom tree into
a triplet $(T_1,T_2,T_3)$ of rooted colored binary blossom trees.
 
These trees differ from the original one in the sense that they are no longer 
automatically well-balanced, that they carry an extra bud attached to the
middle of the root edge, and that the corresponding half-edges connected to their roots       
have three distinct colors $1,2,3$. 
More precisely, we order the triplet in such a way that $T_i$ have a root half-edge of color $i$. 

Conversely given three such rooted colored binary blossom trees $(T_1,T_2,T_3)$,
we attach them by their roots in clockwise cyclic order.
The result is a vertex-marked well-colored binary blossom tree: its three ``unmatched" leaves have
colors $1,2,3$ and we therefore pick that of color $2$ for the root which makes 
it well-balanced and well-colored.

\fig{The pictorial representation of the functional equation 
$V_1=x_1+tV_1V_2+tV_3V_1$ for the generating functions $V_i$ of 
colored blossom trees with a first half-edge of color $i$. The color 
of half-edges are indicated by flags and appear in clockwise (resp.
counterclockwise) order around white (resp. black) vertices.}
{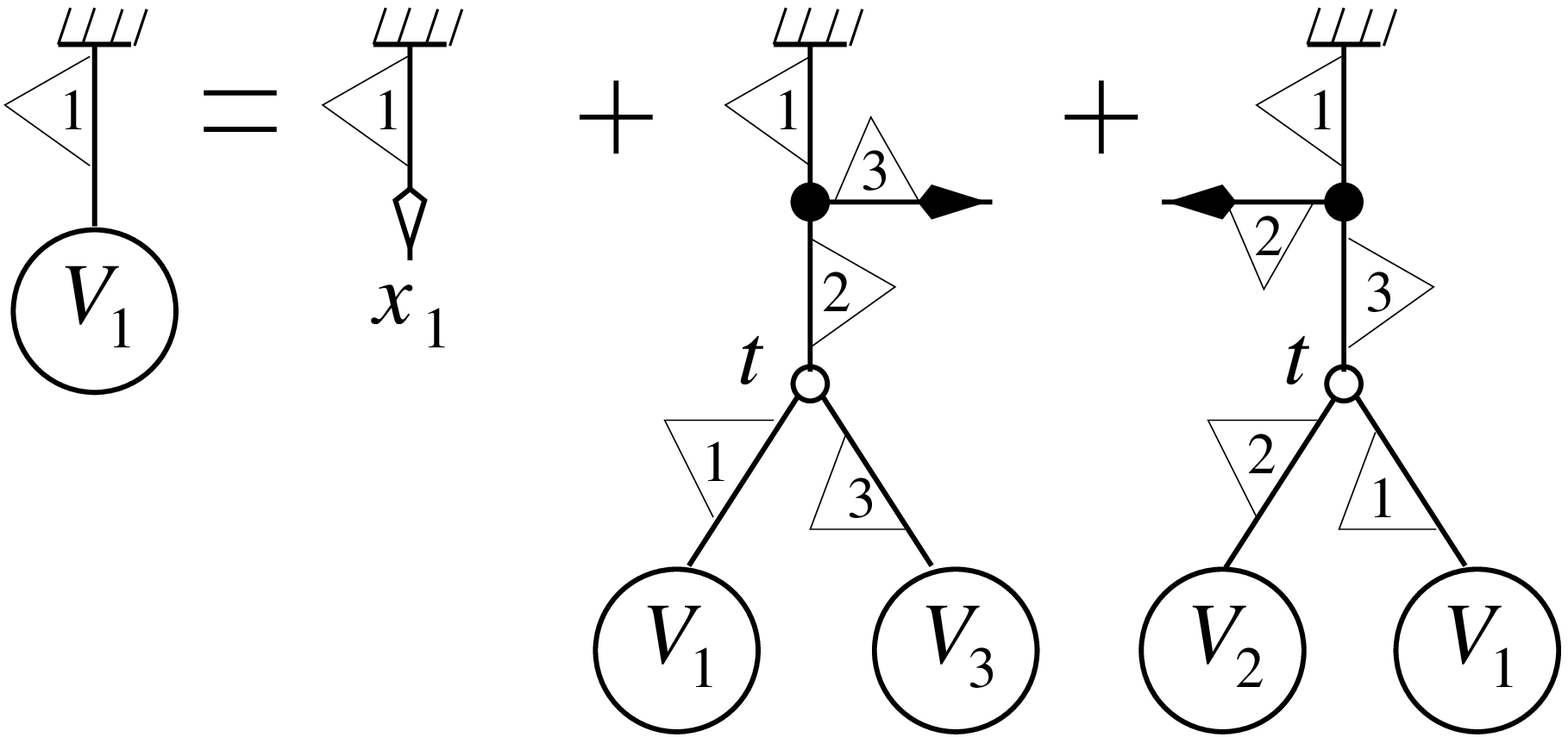}{10.truecm}
\figlabel\threetrees

This bijection translates into $dE/dt=V_1V_2V_3$, where  
$V_i(t;x_1,x_2,x_3)$ denotes the generating function for rooted blossom 
trees with a root half-edge of color $i$, and weights $t$ per white vertex and $x_i$ per leaf
of color $i$ (excluding the root).  We have the relations:
\eqn\relatvi{ \eqalign{V_1&= x_1 + t V_1 V_2 +t V_3 V_1 \cr
V_2&= x_2 + t V_2 V_3 +t V_1 V_2 \cr
V_3&= x_3 + t V_3 V_1 +t V_2 V_3 \cr}}
illustrated in Fig. \threetrees. 
Upon identifying $U_i=t V_i$, we get:
\eqn\derivE{{d\over dt}E(t;x_1,x_2,x_3)={U_1U_2U_3\over t^3}} 
an equation already obtained in Ref. \DEG.

\fig{The quadruplet of trees $T_1$ of root color $1$, $T_2$ of
root color $2$, and $T_3$, $T_3'$ of root color $3$ are fused into
a blossom tree with a marked inner edge by connecting them cyclically 
as shown. The position of the bud follows.}{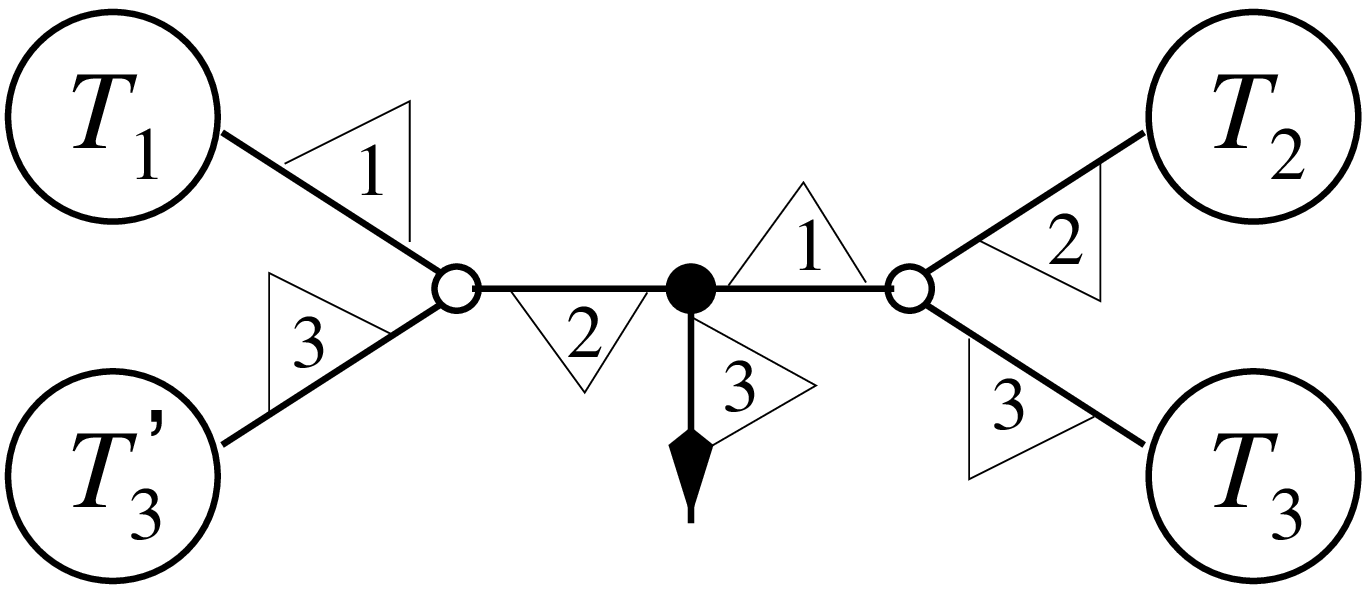}{8.truecm}
\figlabel\fourtrees

Let us now consider what follows from the bijection if instead of 
marking a white vertex we mark an {\it inner edge} of the well-colored blossom 
tree. Noting that there is exactly
one less inner edge than white vertices, the generating function of such objects
simply reads $dE/dt -E/t$.
Now marking an inner edge amounts to actually cutting it out together with its
bud and two adjacent white vertices, thus creating {\it four} colored 
blossom trees of the type generated by the $V_i$'s. In this
quadruplet, exactly two have roots of the same color: they are those
originally separated by the bud in the middle of the marked edge. 

Conversely given a quadruplet of such rooted colored blossom trees
exactly two of which have roots of the same color, there is a unique way
of fusing them into a well-colored well-balanced rooted 
blossom tree with a marked edge, as shown in Fig. \fourtrees.

This translates into the relation
\eqn\transrel{ ({d \over dt} -{1 \over t}) E(t;x_1,x_2,x_3) = {U_1 U_2U_3\over t^3}
(U_1+U_2+U_3) }
which, together with \derivE\ implies the desired formula \lutinit\
for $E$.

\newsec{Discussion and conclusion}

In this paper, we have given a combinatorial proof of the result \lutinit\
for the generating function of vertex-tricolored planar triangulations 
with a marked edge of color (12). This was done by adapting the construction
of Ref. \BMS\ by first adding colors to the various objects involved
and then deriving closed formulas for the generating functions
by suitably marking them.
Let us now discuss a natural generalization of \lutinit\ to other
graph-coloring problems.

The construction of Ref. \BMS\ was carried out for general multivalent
graphs called constellations. Among them are the rooted Eulerian $k$-gonal tessellations,
namely duals of the rooted bipartite $k$-valent planar graphs. The $k$-gonal faces of
the tessellation are naturally bicolored into alternating black and white faces
while its vertices are $k$-colored with 
colors $1,2,\cdots,k$ by demanding that they appear in clockwise (resp.
counterclockwise) cyclic order around white (resp. black) faces, and that the
root edge be of color (12). This coloring is then unique and we wish to compute
the generating function $E(t;x_1,\cdots,x_k)$ for
these objects with a weight $x_i$ per vertex of color $i$ and a weight $t$ per
edge of color (12). For $k=2,3$, this exhausts all possible vertex colorings
of these graphs with $k$ colors. For $k>3$ however, the prescription of cyclic
ordering of colors restricts the configurations to very particular $k$-colorings
of the vertices of those graphs. The absence of entropy of coloring is what
makes the problems simple. 
Repeating the steps of Ref. \BMS\ and adding colors on the way, we arrive at the 
following generalization
\eqn\lutinter{ \eqalign{  
E(t;x_1,...,x_k) & = {U_1U_2...U_k\over t^{k-1}}
\big(1-{U_1U_2...U_k\over t^{k-3}}\sum_{1\leq i<j\leq k} {1\over U_iU_j}\big) 
\cr & U_i  \left(1-t^{3-k}\sum_{j\neq i}{1\over U_j}\prod_{m\neq i}U_m\right) 
= t x_i\cr}}
where $U_i\equiv U_i(t;x_1,\cdots,x_k)$ satisfy $U_i(0;x_1,\cdots,x_k)=0$
and generate the numbers of suitable well-colored well-balanced blossom trees.
Like in the $k=3$ case, we may derive the above relation in two steps.
We first get an explicit expression for $dE/dt$ by
marking a white vertex, thus creating a $k$-uplet of colored blossom trees
with roots of all colors. The correspondence is a bijection, hence
\eqn\derivk{{d\over dt}E(t;x_1,\cdots,x_k)={U_1U_2\cdots U_k\over  t^k}}
Alternatively, marking an inner edge of the well-colored
blossom tree, rather than a white vertex, leads to an expression 
for $dE/dt -E/t$. This marking amounts to erasing the inner edge 
together with its buds and two adjacent
white vertices, thus creating a $2(k-1)$-uple of colored blossom trees. 
Around each white vertex
(with its $(k-1)$ attached trees) exactly one color of root is missing among 
$1,2,...,k$. Overall a pair of colors $i\neq j$ are missing from the two sets of roots.
This translates into the following relation
\eqn\folk{ ({d\over dt} -{1 \over t})E(t;x_1,...,x_k)={1\over t^{2k-3}}
(U_1U_2...U_k)^2\sum_{1\leq i<j\leq k} {1\over U_iU_j} }
which, together with \derivk, implies the formula \lutinter\ for $E$.
When $x_1=x_2=\cdots=x_k=1$, \lutinter\ reduces to the following
equation for the generating function $E(t)$ of rooted Eulerian $k$-gonal
tessellations:
\eqn\redinter{\eqalign{E&={U^k\over t^{k-1}} \left(1-{k(k-1)\over 2}
{U^{k-2}\over t^{k-3}}\right)\cr U&\left(1-(k-1){U^{k-2}\over t^{k-3}}
\right)=t\cr}} 
in agreement with the matrix model result detailed in Appendix A.
The formula \lutinit\ and its generalization
\lutinter\ may be compared to that derived in Ref. 
\BMS\ in the context of constellation enumeration (eqs. 10 and 11).  
The latter involves generating functions $u_m$ for general Eulerian
trees which turn out to be related to our $U_j$'s at $t=1$ through
$u_m=\prod_{j\neq m} U_j$.

\appendix{A}{Enumeration of rooted bipartite $k$-valent planar graphs via matrix model}

Hermitian matrix integrals have been extensively used to generate
(planar) graphs \BIPZ. Here
we wish to generate the dual graphs to planar Eulerian 
$k$-gonal tessellations,
counted with a weight $\sqrt{t}$ per $k$-gon, namely the bipartite 
$k$-valent planar graphs with a weight $\sqrt{t}$ per vertex (or 
equivalently a weight $t$ per white vertex in a natural bicoloration
of the vertices of the bipartite graph).
The case of rooted bicubic maps of Ref. \TUT\ corresponds to $k=3$ in this language.
This is readily obtained by computing the following integral over
two Hermitian matrices $A,B$ of size $N\times N$:
\eqn\matint{\eqalign{ Z_N(t) &= \int dA dB e^{-N\, {\rm Tr}(V(A,B))} \cr
V(A,B)&= {AB} -{\sqrt{t}\over k}(A^k+B^k) \cr}}
where the integration measure is the standard Haar measure, normalized in such a way
that $Z(0)=1$. Indeed, the generating function for rooted objects\foot{The function 
$t{df_0(t)\over dt}$ counts $k$-valent planar graphs with a marked white vertex,
which is $1/k$ of the counting function for $k$-valent planar graphs with a marked edge.} 
is simply $E(t)= kt {df_0(t)\over dt}$
where the ``planar free energy" $f_0(t)$ reads
\eqn\planafren{ f_0(t) =\lim_{N\to \infty} {1\over N^2} {\rm Log} \, Z_N(t) }
Indeed, ${\rm Log}\, Z_N(t)$ is formally computed as a power series expansion of $t$ each term of
which corresponds to the construction of all possible closed connected (fat)graphs 
by connecting pairs of half-edges of vertices of the type $A^k$ and $B^k$ 
into bicolored edges of type $AB$. The result is a connected bicolored $k$-valent graph,
weighted by its inverse symmetry factor (from the sum over all possible edge pairings
and the various combinatorial factors of the expansions), as well as a weight 
$N\sqrt{t}$ per vertex, $1/N$ per edge, and $N$ per face, giving an overall
factor $N^{2-2g}$, $g$ the genus of the surface covered by the graph. Taking
$N\to \infty$ therefore selects only the genus zero (planar) contributions.

The computation of $Z_N(t)$ is performed through a sequence of standard steps
[\xref\DGZ,\xref\EY].
We first reduce the integral \matint\ to one over eigenvalues, by first performing
the change of variables from $A,B$ to their diagonal form and unitary diagonalization  
matrices, and then using the Itzykson-Zuber formula to disentangle the interaction
term $AB$, to finally integrate out the unitary matrices. The result reads
\eqn\resul{ Z_N(t)=\int \prod_{i=1}^N \Delta(\lambda) \Delta(\mu)
d\lambda_i d\mu_i e^{-N V(\lambda_i,\mu_i)} }
where $\Delta(\lambda)=\prod_{1\leq i<j\leq N} (\lambda_i-\lambda_j)$ is the Vandermonde
determinant of the diagonal matrix $\lambda={\rm diag}(\lambda_1,...,\lambda_N)$.
Next we introduce monic bi-orthogonal polynomials w.r.t. the weight
$w(x,y)=e^{-NV(x,y)}$, namely $p_n(x)=x^n+O(x^{n-1})$, $q_m(y)=y^m+O(y^{m-1})$ satisfying
\eqn\orthopol{ (p_n,q_m)\equiv \int dx dy w(x,y) p_n(x) q_m(y) =h_n \delta_{m,n}}
where the formal two-dimensional integration over $x,y$ is defined through the
Feynman rules for matrices of size $1\times 1$, namely \DEG\
\eqn\rulint{ \int dx dy x^a y^b e^{-N xy} \equiv \delta_{a,b} {\Gamma(a+1)\over N^a} }
The symmetry of $V(x,y)=V(y,x)$ allows to identify $p_n(x)=q_n(x)$.
The function $Z_N(t)$ is then simply expressed in terms of the normalization factors
$h_n$ as
\eqn\zcalc{ Z_N(t) =\prod_{i=0}^{N-1} h_i }

We now introduce operators $Q$ and $P$ acting on the basis $p_n$ as respectively
$Qp_n(x)=x p_n(x)$ and $Pp_n(x)=p_n'(x)$. 
Upon formally integrating by parts, we easily get the relation
\eqn\pandq{ (Pp_n,p_m)=N(\partial_xV(x,y)p_n,p_m)=N(p_n,yp_m)-\sqrt{t}(x^{k-1}p_n,p_m) }
Once expressed on $p_n$, this translates into the operator relation
\eqn\oprel{ {P\over N} = Q^\dagger -\sqrt{t} Q^{k-1} }
where the adjoint is defined wrt the bilinear form $(\ ,\ )$ above. We now use the 
$k$-fold symmetry of $V(\omega x,{\bar \omega} y)=V(x,y)$, where $\omega=e^{2i\pi\over k}$,
which translates into one for $p_n$: $p_n(\omega x)=\omega^n p_n(x)$. Together
with \oprel\ and the fact that $P$ lowers the degree by $1$, this implies the following 
action of $Q,Q^\dagger$ on $p_n$:
\eqn\actqp{\eqalign{ 
Qp_n&= p_{n+1} + r_n p_{n-k+1} \cr
Q^\dagger p_n&= {r_{n+k-1}\over v_{n+1} v_{n+2}\cdots v_{n+k-1}} p_{n+k-1} + v_n p_{n-1} \cr} }
where $v_n=h_n/h_{n-1}$. Substituting these into \oprel\ yields the all-genus solution
for $Z_N(t)$. As we are only interested in the planar limit, we take $N\to \infty$ and
note that for $n/N=x$ fixed the sequences $r_n, v_n$ tend to smooth functions of $x$
still denoted $r(x), v(x)$. We may formally write $Q=z+r/z^{k-1}$ and $Q^\dagger=r (z/v)^{k-1}+v/z$
in terms of the shift operator $z$, which commutes with all functions of $x$ in the planar limit.
Moreover we have $P/N= x/z+O(1/z^2)$, and expressing \oprel\ for the coefficient of $z^2$
and $1/z$ we get the two relations
\eqn\tworel{ r/v^{k-1}=\sqrt{t} \qquad x=v-(k-1)r\sqrt{t}}
For $k\geq 3$, 
introducing $\theta=t^{1\over k-2}$, and $V(x)=\theta v(x)$, we finally get
\eqn\finget{ \theta x=V(1-(k-1)V^{k-2})\equiv \varphi(V) }
The function $V$ is the unique solution of \finget\ vanishing at $x=0$.
The formula \zcalc\ immediately gives access to the planar free energy
\eqn\planerg{ f_0(t)= \int_0^1 dx (1-x)\  {\rm Log}\big({V(x)\over \theta x}\big) }
or to the quantity we are after
\eqn\equant{ E(t)= kt {df_0\over dt} =
{k\over k-2} \theta {df_0 \over d\theta}=-{k\over 2(k-2)}+{k\over k-2}
\int_0^1 dx (1-x){\partial_\theta V\over V}}
Changing variables from $x\to V$, and denoting by $W$ the solution of
\eqn\solU{\theta=W(\theta)(1-(k-1)W(\theta)^{k-2}),\qquad  W(0)=0}
we arrive at
\eqn\arrive{\eqalign{ E(t)&=-{k\over 2(k-2)}+{k\over t^2(k-2)}
\int_0^U {dV\over V}\varphi(V)(t-\varphi(V))\cr
&=-{k\over 2(k-2)}+{k\over t^2(k-2)}\Big(t W-{W^2\over 2}-tW^{k-1}+2{k-1\over k}W^k-{k-1\over 2}
W^{2(k-1)}\Big) \cr}}
In the case $k=3$ where $\theta=t$, we simply have to substitute the solution
$W=(1-\sqrt{1-8t})/4$ of \solU\ into the above, to finally recover Tutte's
result \gentut.
In the general case, we may reduce the last expression of \arrive\ to the same
denominator $(k-2)t^2$ and substitute $\theta$ with $W(1-(k-1)W^{k-2})$ in the numerator, 
which after some drastic simplifications finally gives 
\eqn\finlut{ E(t)= {W^k \over t^2} (1-{k(k-1)\over 2} W^{k-2}) }
Finally, writing $U(t)=\theta^{k-3} W(\theta)$, eqns. \solU\ and \finlut\ simply
amount to the formula \redinter. 
Note finally that for $k=2$, eqns. \tworel\ reduce to $v=x/(1-t)$, and \zcalc\
implies
\eqn\lastnotleast{ f_0(t)= \int_0^1 dx (1-x)\  {\rm Log}\big({v(x)\over x}\big)=-{1\over 2}{\rm
Log}(1-t) }
hence $E(t)=2 tdf_0/dt=t/(1-t)$, in agreement with \redinter\ at $k=2$.

\listrefs
\end